\documentclass[superscriptaddress,reprint,amsmath,amssymb,floatfix]{revtex4-2}
\usepackage{graphicx}% Include figure files
\usepackage{color,soul}
\usepackage[normalem]{ulem}
\usepackage{dcolumn}% Align table columns on the decimal point
\usepackage{bm}% bold math
\usepackage{hyperref}% add hypertext capabilities
\usepackage[mathlines]{lineno}% Enable numbering of text and display math
\begin{document}

\preprint{APS/123-QED}

\title{BGO relaxation dynamics probed with heterodyne detected optical transient gratings}
%\thanks{danny.fainozzi@elettra.eu}

\author{Danny Fainozzi}
\affiliation{Elettra-Sincrotrone Trieste, SS 14 km 163,5 in AREA Science Park, 34149, Trieste, Italy.}
 \email{danny.fainozzi@elettra.eu}
\author{Sara Catalini}
\affiliation{European Laboratory for Non-Linear Spectroscopy
Via Nello Carrara, 50019, Sesto Fiorentino, Italy}
\affiliation{Department of Physics and Geology, University of Perugia, Perugia, Italy.}
\author{Renato torre}
\affiliation{European Laboratory for Non-Linear Spectroscopy
Via Nello Carrara, 50019, Sesto Fiorentino, Italy}
\affiliation{Department of Physics, University of Florence, Florence, Italy.}
\author{Claudio Masciovecchio}
\affiliation{Elettra-Sincrotrone Trieste, SS 14 km 163,5 in AREA Science Park, 34149, Trieste, Italy.}
\author{Cristian Svetina}
\affiliation{Instituto Madrileño de Estudios Avanzados en Nanociencia, C/Faraday 9, 28049, Madrid, Spain}


\begin{abstract}
\noindent
We used optical laser pulses to create transient gratings (TGs) with sub-10 $\mu$m spatial periodicity in a Bismuth Germanate (310) $(Bi_4Ge_3O_{12})$ single crystal at room temperature. The TG launches phonon modes, whose dynamics were revealed via forward diffraction of a third, time-delayed, heterodyne-detected optical pulse. Acoustic oscillations have been clearly identified in a time-frequency window not covered by previous spectroscopic studies and their characteristic dynamic parameters have been measured as a function of transferred momenta magnitude and direction.
\end{abstract}

%\keywords{Suggested keywords}%Use show keys class option if keyword
\maketitle

%\tableofcontents

\section{Introduction}

\noindent
\hspace*{3mm} Bismuth Germanate (BGO) stands out as an optically isotropic material boasting a eulitine cubic crystalline structure, a quality that has garnered significant attention due to its versatile electro-optic, electro-mechanical, and scintillation attributes \cite{raymond2000influence,williams1996optical}. The utility of BGO crystals extends across a spectrum of applications. Notably, they have emerged as crucial components in particle scintillation detectors employed in high-energy physics experiments \cite{brunner2017bgo}. Beyond this, BGO crystals have also found application as a medium for holographic data storage \cite{montemezzani1992electro}, a fundamental material in high-resolution positron emission tomography, a host for solid-state lasers when doped with trivalent rare-earth ions \cite{kamada1993electro}, and an electro-optic material enabling the realization of optical voltage, current, and electric power sensors \cite{li2003optical}.

\begin{figure}%[h!]
\centering
\centering\includegraphics[width=\linewidth]{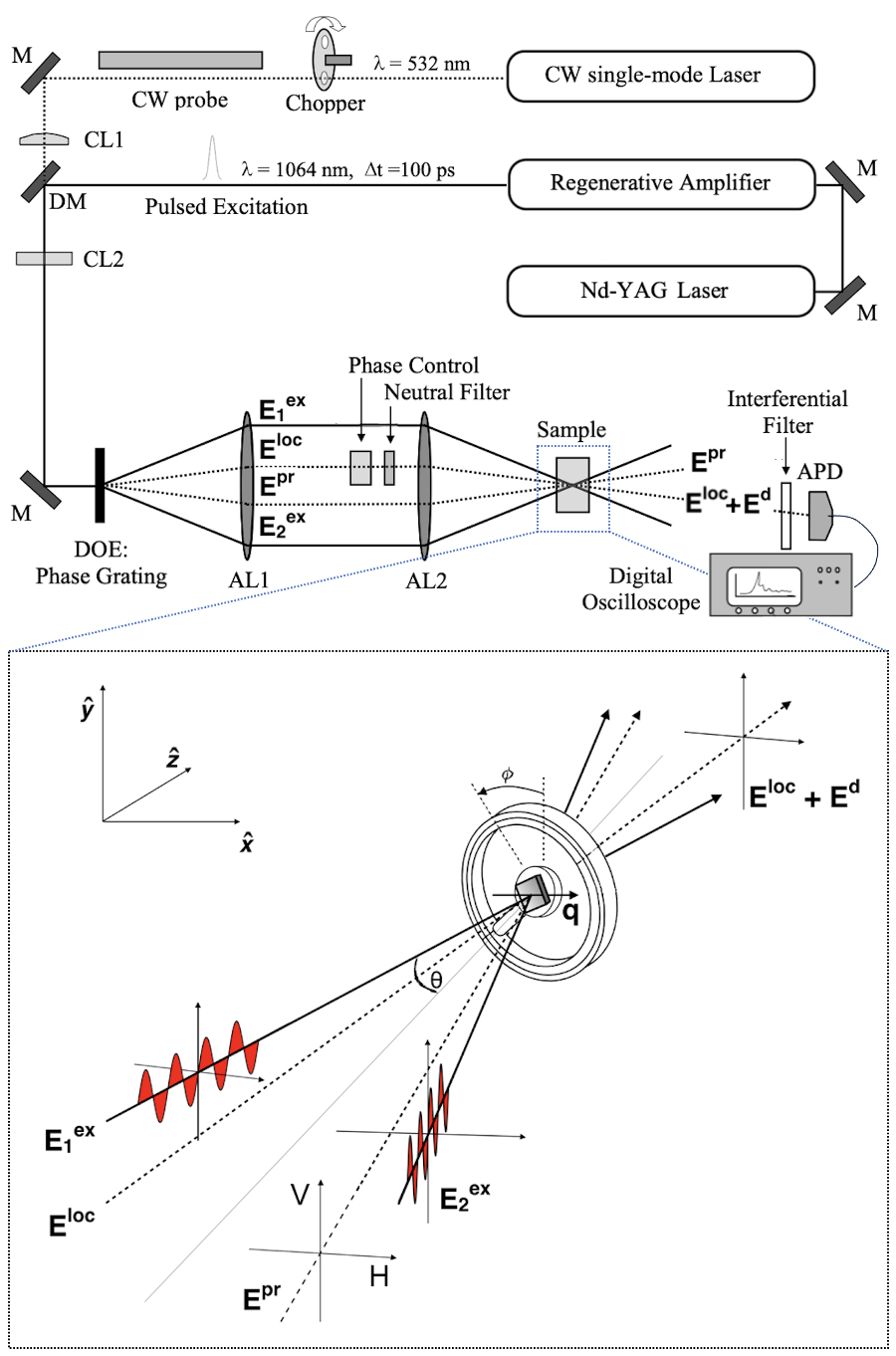}
\caption{Shows the experimental setup, including the laser system. The label M stands for the mirrors, CL1,2 labels the cylindrical lenses, and DM labels a dichroic mirror. DOE is the diffractive optic element, AL1,2 are the achromatic lenses, and APD is an avalanche photodiode. In the zoom-in, we illustrate the polarization configuration and beam directions. $E_1^{\textit{ex}}$ and $E_2^{\textit{ex}}$ are the excitation laser pulses and $E^{\textit{pr}}$ and $E^{\textit{d}}$ are the probing and diffracted beams, respectively. The optical heterodyne detection is obtained with the local field, $E^{loc}$. The sample was mounted on a hollow goniometer, able to rotate on the whole circle angle, \emph{i.e.,} $2\pi$, with a $0.5^{\circ}$ precision.}
\label{fig:fig_1}
\end{figure}
\noindent
\hspace*{3mm} In a TG experiment, temporally coincident pulses of wavelength $\lambda$ (referred to as pumps) are superimposed onto the sample at an intersecting angle of $2\theta$. The interaction between these two pulses leads to the creation of a spatially modulated light intensity pattern, characterized by a period $\Lambda_{\text{TG}}=\lambda/(2\sin \theta)$; as illustrated in Fig. \ref{fig:fig_1}. This structured excitation serves as a transient diffraction grating for a third variably-delayed pulse (the probe), with wavelength $\lambda_{\text{pr}}$, giving rise to a fourth pulse: the diffracted beam (signal).\\
\hspace*{3mm} In this work, the excitation grating is generated by two laser pulses (1064 nm, 25 ps FWHM) obtained by splitting a single pulse through a solid diffraction grating using a solid diffraction grating, their subsequent interference giving rise to an impulsive, spatially periodic alteration in the dielectric constant $\delta\epsilon_{ij}(t)$ within the sample. This spatial variation is characterised by a wavevector $|\vec{q}| = |\vec{k}_1-\vec{k}_2| = \frac{4\pi}{\lambda} \sin\theta$, where $\vec{k}_{1,2}$ represent the wavevectors of each respective pump pulse. A second laser beam (532 nm, continuous wave (CW)), the probe, is directed onto the induced grating at the Bragg angle. This arrangement leads to the emergence of a diffracted beam, spatially differentiated from both the pump and probe beams, yielding the dynamic information from the relaxing grating. The complete description of TG experiments Comprehensive details on Transient Grating (TG) experiments can be found in a multitude of review papers, including references such as \cite{eichler2013laser,torre2007time,clark1989time}. When heterodyne detection is employed \cite{torre2001acoustic,goodno1998ultrafast,maznev1998optical}, the transient grating diffracted field is combined with a reference beam at the detector. This superposition encompasses the local field, which entails an electric field sharing the same frequency as the probe beam and being in phase with it. The resultant signal assumes a general form:
\begin{equation}
S^{\text{HD}}(\vec{q},t) \propto \delta\epsilon_{ij}(\vec{q},t) \propto R_{ijkl}(\vec{q},t).
\end{equation}
Here, the Cartesian indices $i$ and $j$ delineate the polarizations of the local and probe fields, while $k$ and $l$ denote the polarizations of the pump fields. The quantity $\delta\epsilon_{ij}$ represents the change induced in the dielectric constant due to the pump laser, and $R_{ijkl}$ defines the system's response function, governing the dynamic characteristics of the measured experimental quantities. Both $\delta\epsilon_{ij}$ and $R_{ijkl}$ are spatial Fourier components corresponding to the exchanged wave vector $\vec{q}$, $\emph{i.e.,}$ the grating wave vector. Consequently, the signal acquired through heterodyne-detected Transient Grating (HD-TG) directly and linearly captures the relaxation processes defined by the tensor components of the response function. By strategically selecting suitable electric field polarizations, it becomes possible to isolate a specific component of the tensor. It is noteworthy that distinct response components typically correspond to dissimilar relaxation processes \cite{taschin2001translation,glorieux2002thermal,califano2003recent,di2003nonequilibrium,di2003structural,pick2004heterodyne}, wherein both translational and rotational degrees of freedom come into play. The influence of rotational dynamics can be experimentally explored through a comparative analysis of the HD-TG signals $S^{\text{HD}}_{\textit{VVVV}}$, $S^{\text{HD}}_{\textit{HHVV}}$, and $S^{\text{HD}}_{\textit{VVHH}}$ wherein H and V denote horizontal and vertical polarizations, respectively.\\
In the present experimental setup, all the aforementioned polarizations are oriented vertically, thus enabling the focused examination of the individual tensorial component of the response function, specifically $R_{\textit{VVVV}}$. Considering the entities in figure \ref{fig:fig_1}, the resulting forward signal can be expressed as:
\begin{equation}\label{eq:all_signal}
\begin{split}
S(q,t)&\propto \langle|E^d(q,t)+E^{loc}|^2\rangle = \\
&=\langle|E^d(q,t)|^2\rangle + \langle|E^{loc}|^2\rangle +\\
&\hspace{.4cm}+ 2 \langle|E^d(q,t)|\rangle  \langle|E^{loc}|\rangle \cos(\Delta \phi).
\end{split}
\end{equation}
Here, $\Delta\phi$ represents the phase discrepancy existing between the diffracted field $E^d$ and the local field $E^{loc}$, while $\langle\cdot\rangle$ denotes the temporal averaging conducted over an optical period. The trio of terms present on the right-hand side of Eq. \ref{eq:all_signal}, corresponds, respectively, to the homodyne, local field, and heterodyne contributions, as read from left to right. To specifically isolate the latter term, a subtraction operation was executed between two signals characterized by differing phases. This process involved the initial acquisition of a signal denoted as $S_+$, with a corresponding phase $\Delta\phi_+ = 2n\pi$ ($n\in \mathbb{N}$), followed by the acquisition of a secondary signal, labelled as $S_-$, characterized by a phase $\Delta\phi_- = (2n+1)\pi$. This procedure yields the following expression:
\begin{equation}
S^{\text{HD}}(q,t)=[S_+ + S_-]\propto \langle|E^d(q,t)|\rangle  \langle|E^{loc}|\rangle
\label{eq:eq3}
\end{equation}
There are two significant advantages associated with using heterodyne detection over homodyne detection. Firstly, it substantially enhances the signal-to-noise ratio across the entire time range, attributed both to the augmentation of the signal itself and to the elimination of all non-phase-sensitive spurious signals. Secondly, it elevates sensitivity levels, as the recorded signal becomes directly proportional to $\langle|E^d(q,t)|\rangle$, as opposed to being proportional to its square.\\
\section{Experimental Setup}
\begin{figure}%[h!]
\centering
\centering\includegraphics[width=\linewidth]{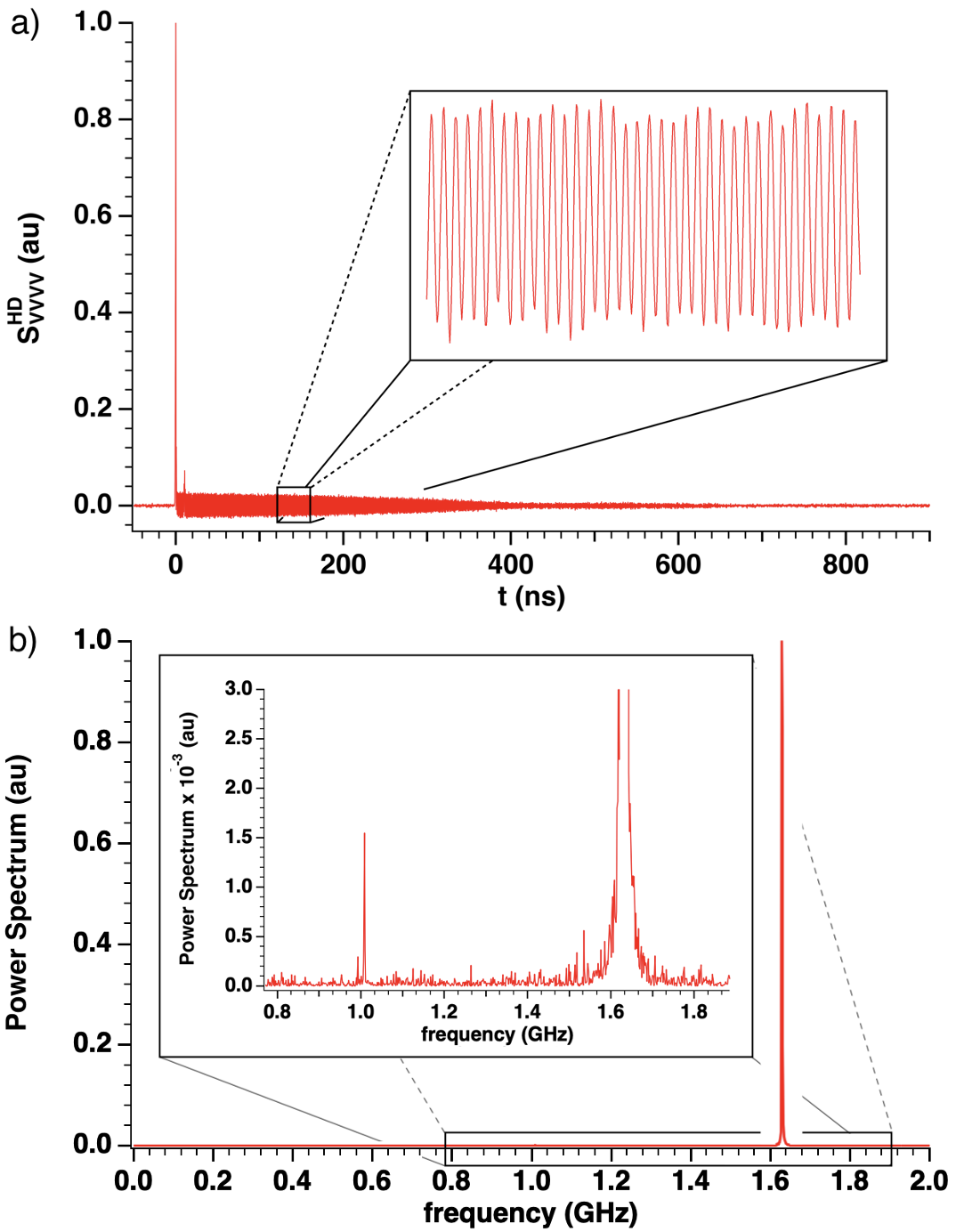}
\caption{Panel a) shows the long trace $S^{\text{HD}}_{\textit{VVVV}}$ acquired up to 900 ns. In the insert, a zoom-in underlying the oscillation is given by acoustic phonons. In panel b), the FFT of the $S^{\text{HD}}_{\textit{VVVV}}$ signal, shows two main frequencies labelled as the LA and TA phonons.}\label{fig:fig_2}
\end{figure}
The BGO sample was sectioned along the (310) crystallographic plane, exhibiting an azimuthal (in-plane) angle of approximately $45^{\circ}$ in relation to the [100] crystallographic direction when the goniometer reading was $\vartheta=0$. The sample's dimensions were 10 $\times$ 10 $\times$ 1 mm$^3$ ($L\times W\times T$). The optical configuration, depicted in Fig. \ref{fig:fig_1} incorporates a diffractive optical element (DOE) serving as a phase grating. By regulating the groove depth and spacing, it becomes feasible to achieve notably high diffraction efficiency, surpassing 80\% for the first two diffraction orders. However, since the experimental setup must manage the diffraction of both pump and probe lasers (1064 nm and 532 nm, respectively) utilizing a single grating, a compromise must be struck. The chosen DOE results in a 12\% diffraction efficiency for the first order at 532 nm and a 38\% diffraction efficiency for the first order at 1064 nm. Six gratings, each with distinct spacing, were employed to manipulate the $\vec{q}$ vector. Assisted by a dichroic mirror (DM), the excitation and probe beams are directed collinearly onto the DOE, which subsequently generates the two excitation pulses ($E_{1,2}^{\text{ex}}$), the probe ($E^{\text{pr}}$), and the reference beam ($E^{\text{loc}}$). These beams are gathered by a primary achromatic lens (AL1), refined through a spatial mask that obstructs other diffracted orders, and then recombined and focused onto the sample using a second lens (AL2). The local laser field is attenuated by a neutral density filter, and its phase is adjusted by passing through a pair of properly etched quartz slabs. The excitation grating formed on the sample mirrors the illuminated DOE phase pattern. In scenarios where AL1 and AL2 share identical focal lengths, the excitation grating exhibits half the spacing of the DOE \cite{maznev1998optical}. This setup inherently fulfils the Bragg condition for all beams, resulting in a remarkably stable phase alignment between the probe and reference beams, a pivotal requisite for implementing heterodyne detection. A cylindrical lens (CL2) is employed to focus the excitation beam onto the DOE. Consequently, the resulting grating generated on the sample extends along the $\vec{q}$ direction for approximately 5 mm. Conversely, the probe beam is focused into a circular spot measuring 0.5 mm on the sample, achieved through the use of lenses CL1 and CL2. During this measurement, laser energy on the sample was minimized to mitigate unintended thermal effects, while the continuous wave beams were gated within a window of approximately 1 ms every 10 ms, facilitated by a mechanical chopper synchronized with the excitation pulses. The average excitation energy amounted to 7 mW (equivalent to 35 $\mu$J per pulse at 100 Hz), while the probing energy was set to 6 mW. The reference beam's intensity is deliberately kept low and is experimentally adjusted, employing a variable neutral filter, to be roughly 100 times smaller than the intensity of the diffracted signal. Given these intensity levels, the experiment operates comfortably within the linear response regime, with no discernible influence of the beam intensities on the HD-TG signal shape. The HD-TG signal, post optical filtration, is measured and captured by the New Focus DC-Coupled Photoreceiver (with a bandwidth of 12 GHz), interfaced with a Tektronix oscilloscope possessing a limiting bandwidth of 7 GHz and a sampling rate of 20 Gs/s. The BGO crystal is mounted on a hollow goniometer with a precision of $0.5^{\circ}$, enabling the sample's rotation within the $\hat{x}-\hat{y}$ plane.
\section{Results}
The signal obtained from the BGO sample was recorded within the time range of 0-1 $\mu$s, utilizing a time step of 50 ps. Each data point is an average derived from 5000 individual recordings, ensuring an excellent signal-to-noise ratio. To eliminate the homodyne contribution and mitigate other undesired influences while markedly enhancing data quality \cite{torre2001acoustic}, each measurement was conducted at two distinct phases $\phi$ of the local oscillator field, separated by $\Delta\phi=\pi$, following Eq. \ref{eq:eq3}. We systematically examined the relaxation process of BGO with respect to the rotation angle $\vartheta$ across six different $|\vec{q}|$ values: $|\vec{q}|=0.6283,1.0134,1.3674,1.7915,2.0854, \text{and } 2.4961$ $\mu$m$^{-1}$. The determination of these wave vectors is influenced by the experimental geometry, introducing a degree of error \textit{i.e.,} approximately 0.8\% for the first two values, around 0.6\% for the intermediary two, and approximately 0.4\% for the final two values. For each distinct wave vector, data was acquired as a function of the angle $\vartheta$ in increments of 45 degrees across a complete circular range. For the case of $|\vec{q}|=2.4961$ $\mu$m$^{-1}$ measurements were also conducted with a 15-degree interval during the first half of the circular range. Fig. \ref{fig:fig_2}a) showcases illustrative HD-TG data, visually revealing that the density dynamics of BGO are characterized by two predominant dynamic processes: an immediate response to the abrupt heating of the sample, succeeded by the subsequent local temperature relaxation attributed to thermal diffusivity. Additionally, oscillations corresponding to phonons launched by the abrupt density fluctuation are evident. Consequently, the measured response function, directly proportional to the $\vec{q}$ component of the induced density variations, can be elucidated through a simple physical model:
\begin{equation}
\label{eq:eq_44}
\begin{split}
S^{\text{HD}}(q,t) =& \frac{1}{2}\Bigl[1+\text{erf}\Bigl(\frac{t}{\sigma}\Bigr)\Bigr] \cdot \Bigl[A\, e^{-\frac{t}{\tau}} + \\ & - \sum_{i} A_i\sin(2\pi\,\nu_i(q)\,t + \phi_i)\, e^{-\frac{t}{\tau_i}}\Bigl].
\end{split}
\end{equation}
\noindent

\begin{figure}%[h]
\centering
\includegraphics[width=8.5cm,keepaspectratio]{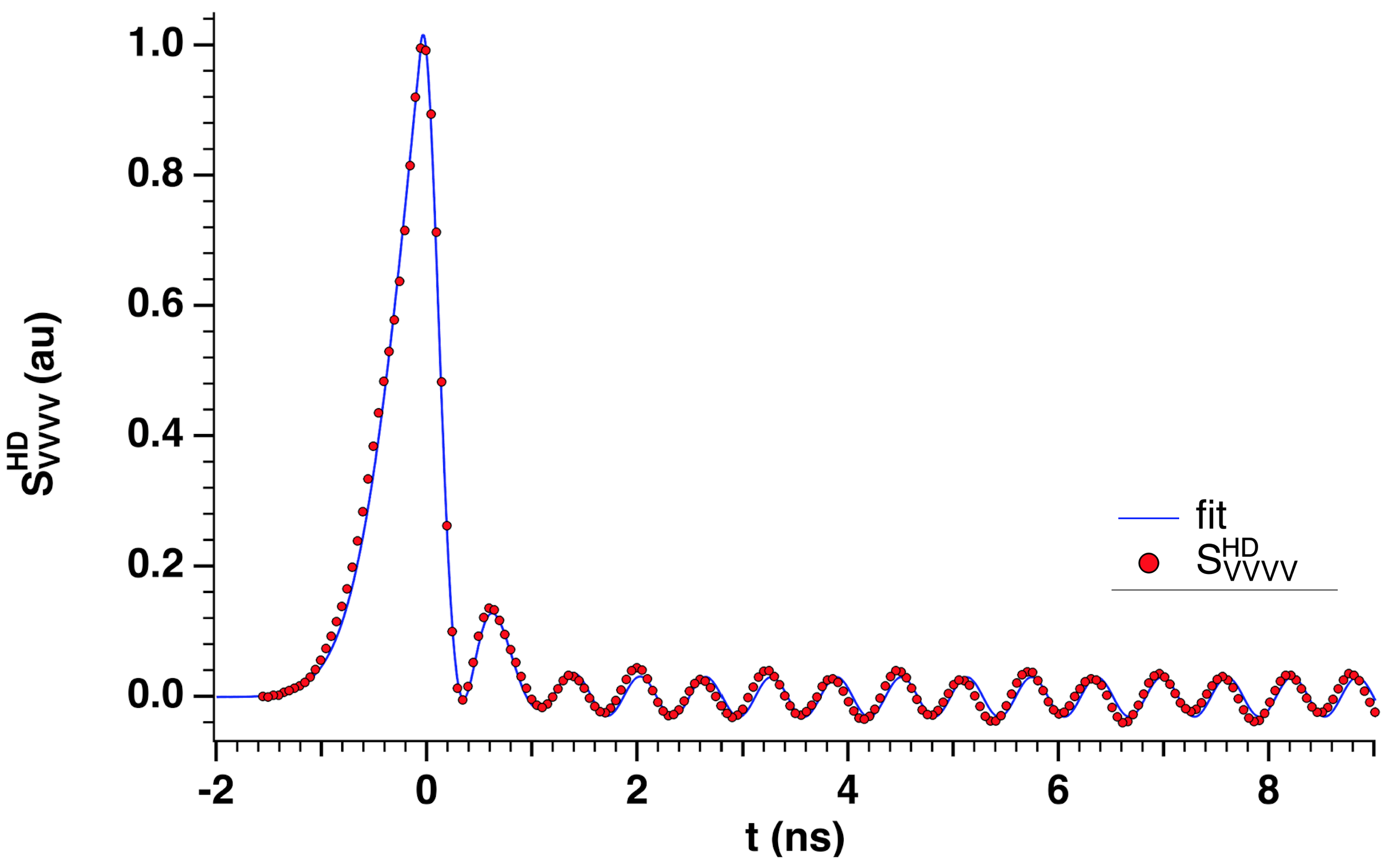}
\caption{The first few ns of the $S^{\text{HD}}$ signal and its fit obtained using Eq. \ref{eq:eq_44}, excluding the laser spurious contribution.}
\label{fig:fig_3}
\end{figure}
\begin{figure}%[h]
\centering
\centering\includegraphics[width=\linewidth]{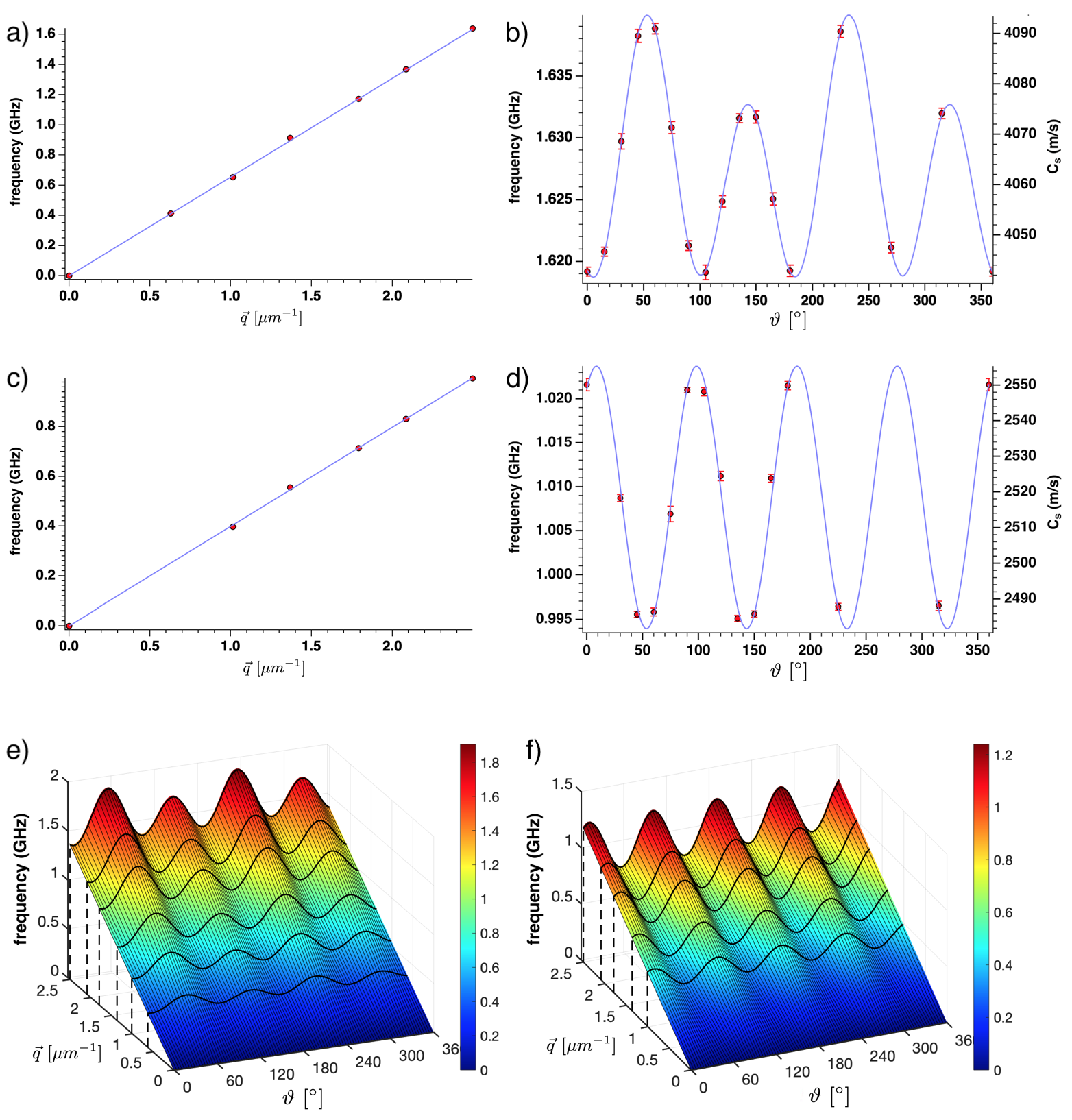}
\caption{Panels a) and b) showcase the linear and angular dispersion profiles of the LA phonon at a $|\vec{q}|$ value of 2.4961 $\mu$m$^{-1}$. Similarly, panels c) and d) exhibit the dispersion patterns of the TA phonon, corresponding to the same value of $|\vec{q}|$. Panels e) and f) provide a 3D reconstruction detailing the linear and angular dispersion characteristics of the LA and TA phonons, respectively. For enhanced clarity, the oscillations have been magnified by factors of 25 (LA) and 15 (TA) around their respective mean values for a given momentum. Notably, black lines denote fitted curves applied to the angular dispersion data across various momentum values.}
\label{fig:fig_4}
\end{figure}

\noindent
Here, the \textit{erf} function encapsulates the sudden rise of the signal (with $\sigma$ representing the rise width), succeeded by an exponential decay characterized by a time constant $\tau$ and additional $i$ attenuated vibrational modes; outcomes are depicted as blue lines in Figs. \ref{fig:fig_3}. Following this, Fourier Transforms (FTs) were performed on the disparities between the measured traces and their corresponding exponential fits, as illustrated in Fig. \ref{fig:fig_2}b). The resulting FTs of the waveforms reveal the distinct presence of two well-defined modes, corresponding respectively to the Transverse Acoustic phonon (TA) and the Longitudinal Acoustic phonon (LA). The findings concerning the frequencies of these oscillations are portrayed in Fig \ref{fig:fig_4}.\\
\hspace*{3mm} Throughout the data acquisition process, the stability of the Nd-YAG laser exhibited inconsistency, resulting in the presence of several harmonics of the laser within the signal. The amplitude of these harmonics varied irregularly within the same scan and consequently from one scan to another. The most prominent contribution from the laser was the harmonic peak at 0.767 GHz, which was duly accounted for during the fitting procedure. Unfortunately, due to the aforementioned instability, a comprehensive fitting of the signal (depicted in Fig. \ref{fig:fig_2}a))was not feasible. Instead, fitting was confined to the initial ns range (as shown in Fig. \ref{fig:fig_3}), during which the laser's contribution could be considered relatively constant. Consequently, the assessment of the decay time for acoustic oscillations was not possible, leading to the assumption of undamped sinusoidal functions during the fitting process. Moreover, the same instabilities posed challenges in obtaining a reliable value for the thermal decay constant $\tau$. Fig. \ref{fig:fig_2}b) displays the normalized FT of the representative HD-TG BGO data presented in Fig. \ref{fig:fig_2}a). Within this plot, a distinct peak emerges at 1.63 GHz, signifying the longitudinal acoustic (LA) phonon. In the inset of the figure, a closer view reveals a fainter trace of a transverse acoustic (TA) phonon around 1 GHz. The presence of the TA phonon can be attributed to the sample being cut along the (310) crystallographic plane, introducing a small component of transferred momentum $\vec{q}$ in the \textit{$\hat{z}$} plane. To precisely deduce the frequencies of the LA and TA phonons, FT fitting was performed using Gaussian functions to pinpoint the peak positions accurately. The associated error for each frequency value was determined based on the standard deviation of the Gaussian curve.\\
\hspace*{3mm} In Fig. \ref{fig:fig_4}a), the linear dispersion of the LA phonon is presented, measured at $\vartheta = 45^{\circ}$. The angular scale was established during the experiment to align with $\vartheta = 0^{\circ}$ when a specific side of the sample was oriented parallel to the ground. This alignment was expected to coincide with one of the crystallographic directions of the BGO crystal, potentially leading to a maximum or minimum in the rotational dispersion frequency. In Fig. \ref{fig:fig_4}b), the rotational dispersion of the LA phonon is depicted, measured at $|\vec{q}|=2.4961$ $\mu$m$^{-1}$ alongside the corresponding sound propagation speed within the material. Figs. \ref{fig:fig_4}c) and \ref{fig:fig_4}d) extend this presentation to encompass the corresponding linear and rotational dispersions for the TA phonon. The rotational dispersion of the LA phonon exhibits a periodicity of $\pi$, prompting the application of a double sine function for fitting. Conversely, the rotational dispersion of the TA phonon displays a periodicity of $\pi/2$, lending itself to a single sine function fitting. Each rotational angle yields a linear fit, while each transferred momentum yields sinusoidal fits, collectively contributing to the formulation of the 3D dispersions shown in Figs. \ref{fig:fig_4}e) and \ref{fig:fig_4}f). Notably, the sinusoidal amplitudes of the fits, visualized as black lines in the figures, are magnified by factors of 25 and 15 for the LA and TA phonons, respectively, around their respective mean values for a given momentum. With the parameters derived from the fitting, the 3D dispersions can be effectively modelled as follows:
\begin{equation}\label{eq:3D_TA}
\begin{split}
LA(|\vec{q}|,\vartheta)&=A|\vec{q}|+B|\vec{q}|\sin(f_1\vartheta+\varphi_1)+\\
&+C|\vec{q}|\sin(f_2\vartheta+\varphi_2)
\end{split}
\end{equation}
\begin{equation}\label{eq:3D_TA_2}
TA(|\vec{q}|,\vartheta)=A|\vec{q}|+B|\vec{q}|\sin(f_1\vartheta+\varphi_1)
\end{equation}
The values of the parameters are shown in table \ref{tab:tab2}.

\begin{center}
\begin{table}[h!]
\resizebox{0.4\textwidth}{!}{%
\begin{tabular}{c|cc} 
 & $LA(|\vec{q}|,\vartheta)$ & $TA(|\vec{q}|,\vartheta)$ \\
\hline\hline
$A$ & $0.6521\pm 0.0002$ & $0.4041\pm 0.0001$  \\
$B$ & $-0.0035\pm 0.0004$ & $0.0060\pm 0.0002$ \\
$C$ & $0.0015\pm 0.0004$ & / \\
$f_1$ & $0.0701\pm 0.0006$ & $0.0700\pm 0.0002$ \\
$f_2$ & $0.0351\pm 0.0008$ & / \\
$\varphi_1$ & $0.99\pm 0.08$ & $0.96\pm 0.02$ \\
$\varphi_2$ & $-0.31\pm 0.14$ & / \\
\hline
\end{tabular}}
\caption{The values of the parameters obtained from the functions in Eqs. \ref{eq:3D_TA} and \ref{eq:3D_TA_2}, used to fit the 3D dispersion of the LA phonon and TA phonon, respectively.}\label{tab:tab2}
\end{table}
\end{center}

\section{Conclusions}
We collected optical Transient Grating signal studying the propagation of the high-frequency acoustic waves in Bismuth Germanate $Bi_4Ge_3O_{12}$ (310), in dependence on the transferred momentum and the rotation angle of the sample. By exploiting the heterodyne detection of the employed setup, we were able to produce data with excellent S/N ratios, compatible with values obtained with X-Ray TG spectroscopy \cite{rouxel2021hard}. Further investigations will be conducted to study thermal and acoustic decay mechanisms.

\begin{acknowledgments}
\noindent
The authors thank Dr. P. Bartolini for providing continuous assistance in setting up of the electronics.
\end{acknowledgments}

% whether or not they are actually referenced in the text. This is an appropriate
% for the sample file to show the different styles of references, but the authors
%\nocite{*}

\bibliography{BGO_article}% Produces the bibliography via BibTeX.

\end{document}